\documentclass[structabstract]{aa}  
\usepackage{graphicx}
\usepackage{epstopdf}
\usepackage{epsfig}
\usepackage{color}
\usepackage{multirow}
\usepackage{rotating}
\usepackage{lscape}
\usepackage{longtable}
\usepackage{lipsum} 
\usepackage{tablefootnote}
\usepackage{txfonts}
\begin{document}
   \title{New Insights into Time Series Analysis}
   \subtitle{I - Correlated observations}

   \author{C. E. Ferreira Lopes$^{1}$ \and N. J. G.~Cross$^1$}
   \institute{
          SUPA (Scottish Universities Physics Alliance) Wide-Field Astronomy Unit, Institute for Astronomy, School of Physics and Astronomy, University of Edinburgh, Royal Observatory, Blackford Hill, Edinburgh EH9 3HJ, UK \\ \email{cfl@roe.ac.uk}  
             }

   \date{Received xxx, 2015; accepted xxx, 2015}

 
  \abstract
   {The first step when investigating time varying data is the detection of any reliable changes in star brightness. This step is crucial to decreasing the processing time by reducing the number of sources processed in later, slower steps. Variability indices and their combinations have been used to identify variability patterns and to select non-stochastic variations, but the separation of true variables is hindered because of wavelength-correlated systematics of instrumental and atmospheric origin, or due to possible data 
   reduction anomalies. }
   {The main aim is to review the current inventory of correlation variability indices and measure the efficiency for selecting non-stochastic variations in photometric data.}
   { We test new and standard data-mining methods for correlated data using public time-domain data from the WFCAM Science Archive (WSA). This  archive contains multi-wavelength calibration data (WFCAMCAL) for $\rm
   216,722$ point sources, with at least 10 unflagged epochs in any of five filters (YZJHK), which were used to test the different indices against. We improve the panchromatic variability indices and introduce a new set of variability indices for preselecting variable star candidates. Using the WFCAMCAL Variable Star Catalogue (WVSC1) we delimit the efficiency of each variability index. Moreover we test new insights about these indices to improve the efficiency of detection of time-series data dominated by correlated variations.}
   { We propose five new variability indices which display a high efficiency for the detection of variable stars. We determine the best way to select 
   variable stars using these and the current tool inventory. In addition, we propose an universal analytical expression to select likely variables using
   the fraction-of-fluctuations on these indices ($f_{\rm fluc}$). The $f_{\rm
   fluc}$ can be used as an universal way to analyse photometric data since it
   displays a only weak dependency with the instrument properties. The  variability indices computed in this new approach allow us to reduce misclassification and these will be implemented in an automatic classifier  which will be addressed in a forthcoming paper in this series. }


   \maketitle


\section{Introduction}	
	
The tremendous development in astronomical instrumentation and automation during
the last few decades has given rise to several questions about how to analyse
and  synthesize the growing amount of data. Recently, various dedicated
telescope  systems, both on the ground and in space, have been used for
wide-field shallow, low resolution, multi-epoch, imaging surveys, scanning  the
sky in different wavebands with aims ranging from comprehensive stellar
variability  searches to exoplanet hunting e.g. (PanSTARRS,
\cite{Kaiser-2002};OGLE,  \cite{Udalski-2003}; SUPERWASP, \cite{Pollacco-2006}; 
CoRoT, \cite{Baglin-2007}; NSVS, \cite{Hoffman-2009}; Kepler,
\cite{Borucki-2010}). These data have led to many discoveries in several areas 
of modern astronomy: asteroseismology, exoplanets and stellar evolution 
\citep[e.g.,][]{Huber-2012,De-Medeiros-2013, Walkowicz-2013, Paz-Chinchon-2015}.
The next generation of these surveys, such as Gaia \citep[][]{Bailer-Jones-2013}
and the VISTA Variables in V\'ia L\'actea survey (VVV;
\citealt[][]{Minniti-2010}), are providing a high data flow for a wide range of 
science applications in order to understand the dynamics and stellar variability of the Milky Way galaxy.

The first step to investigating time varying data is the detection of any reliable changes in star brightness \citep[e.g.][]{Welch-1993,Stetson-1996,Wozniak-2000, Shin-2009,Ferreira-Lopes-2015}. This step is crucial to decreasing the running time by reducing the number of sources that slower steps, such as period finding and classification, are run on. The stochastic variations are mainly related to very bright sources, caused by saturation of the detector whereby the flux within the aperture will bleed out into nearby pixels and the measured magnitude becomes dependent on the sky brightness and seeing, or very faint sources where the sky noise dominates, providing an increase in the uncertainty of the measurements, and a dependency on the sky brightness and seeing. Variability indices and their combinations have been used to identify variability patterns and to select  non-stochastic variations \citep[e.g.][]{Damerdji-2007,Shin-2009,Ferreira-Lopes-2015}, but the separation of true variables from noisy data is hindered because of wavelength-correlated systematics of instrumental and atmospheric origin, or due to possible data reduction anomalies. Detection  methods have been optimized for specific variability signals to detect supernovae, microlensing, transits, and other variable sources \citep[e.g.][]{Alard-1998,Wozniak-2000,Gossl-2002,Becker-2004,Corwin-2006,Yuan-2008,Renner-2008}. An important step to optimising this process is to review the current inventory of variability indices and determine the efficiency level for selecting non-stochastic variations in photometric data.

\begin{table*}[htbp]
 \caption[]{Variability Indices analyses in the present work. The  description
 of terms used in this indices are discriminate in Sect.~\ref{ourind} and 
 \ref{improv_ind}.}\label{tvar01}
 {\centering    
 \begin{tabular}{c c c }        
 \hline\hline                 
 Index & Definition & Reference \\    
 \hline                        

    $I_{WS}$ & $   \sqrt{\frac{1}{n\cdot(n-1)}}\sum_{n=1}^{N-1}   \left( \frac{ x_n-\mu }{e_n} \right ) \cdot \left(  \frac{ x_{n+1}-\mu }{e_{n+1}}  \right )  $ & \citealt[][]{Welch-1993} \vspace{0.25cm}  \\ 
    $J_{WS}$ & $ \sum_{n=1}^{N-1} sign(\delta_{n} \delta_{n+1}) \sqrt{\left| \delta_{n} \delta_{n+1} \right|}  $ & \citealt[][]{Stetson-1996} \vspace{0.25cm} \\ 
    $K_{WS}$ & $\frac{1/N\sum_{i=1}^{N} \left| \delta_{i}\right|  }{\sqrt{1/N\sum_{i=1}^{N}  \delta_{i}^{2}}}$ & \citealt[][]{Stetson-1996} \vspace{0.25cm} \\ 
    $L_{WS}$ & $  \left( J_{WS}\cdot K_{WS} \right)/0.789  $ & \citealt[][]{Stetson-1996} \vspace{0.25cm}  \\    
    $I_{\rm pfc}^{(s)'}$ & $\! \frac{1}{n_{s}}\sum_{i=1}^{n}\left [  \sum_{j_1=1}^{m-(s-1)}\!\!\!\cdots \left ( \sum_{j_s=j_{(s-1)}+1}^{m} \!\!\! \Lambda_{ij_1\cdots j_{s}}^{(s)}  \sqrt[s]{\left| \Gamma u_{ij_1} \cdots \Gamma u_{ij_s}  \right|} \right ) \right ]$ & \citealt[][]{Ferreira-Lopes-2015}$^{1}$ \vspace{0.25cm}  \\    
    $I_{\rm fi}^{(s)}$ & $0.5 \cdot \left \{ 1+\frac{1}{n_{s}}\sum_{i=1}^{n}\left [  \sum_{j_1=1}^{m-(s-1)}\cdots \left ( \sum_{j_s=j_{(s-1)+1}}^m \Lambda_{ij_1\cdots j_{s}}^{(s)} \right ) \right ]  \right \}$ & \citealt[][]{Ferreira-Lopes-2015} \vspace{0.25cm}  \\   
    $K_{fi}^{(s)}$ & $\frac{N_{s}^{+}}{N_{s}}$ & the present work  \vspace{0.25cm}\\   
    $L_{pfc}^{(s)}$ & $\frac{1}{N_{s}}\sum_{k=1}^{N_{box}} Q_{(s,k)}$ & the present work  \vspace{0.25cm}  \vspace{0.25cm} \\ 
    $M_{pfc}^{(s)}$ & $med[ Q_{(s)} ]$ & the present work  \vspace{0.25cm}  \vspace{0.25cm} \\ 
    $FL^{(s)}$ & $F^{(s)}\times L_{pfc}^{(s)}$ & the present work  \vspace{0.25cm}  \vspace{0.25cm} \\ 
    $FM^{(s)}$ & $F^{(s)}\times M_{pfc}^{(s)}$  & the present work  \vspace{0.25cm}  \\

 \hline                                   
 \end{tabular}\\ }
     \scriptsize
     \vspace{0.1in}
     \hspace{2cm} {\bf 1. } Unfortunately the first version of $I_{pfc}^{(s)'}$ indices was incorrectly defined. Therefore, the authors have since added an erratum with the correct form..
  \label{tabfits}
\end{table*}

The second step is to determine the main periods. There are various methods used in astronomy for frequency analysis, to name a few: the Deeming method \citep[][]{Deeming-1975}, PDM-Jurkevich \citep[][]{Stellingwerf-1978, Dupuy-1985}, string length minimization \citep[][]{Lafler-1965,Stetson-1996,Clarke-2002}, information entropy \citep[][]{Cincotta-1995}, the analysis of variance \citep[ANOVA,][]{Schwarzenberg-Czerny-1996} and the Lomb-Scargle and its extension using error bars \citep[][]{Lomb-1976,Scargle-1982,Zechmeister-2009}. These methods are based on the fact that the phase diagram of the light curves (LCs) is smoothest when it is visualized using its real frequencies. Assessment of the significance of these frequencies is a pertinent  problem due to non-Gaussianity, multi-periodicity, non-periodic variations, and the manner of how they should be taken into account \citep[][]{Suveges-2014}. From this view the variability indices are a fundamental part of the variability  analysis in order to save running time and decrease the number of miscalculations in the frequency analysis. The detection of non-periodic variables, transients, and other aspects in regard to the significance of peaks in a periodogram has not been completely solved yet.

The last point is that the variability classification is intrinsically related with the determination of reliable periods and determining a set of parameters 
that allows us to distinguish all variability types. Automatic classifiers based on machine learning have been applied to several large time-series datasets \citep[e.g.][]{Wozniak-2004,Debosscher-2007,Sarro-2009,Blomme-2010,Richards-2011,Dubath-2012}. The inclusion of periodic and non-periodic features, statistics and more sophisticated model parameters have improved automatic classifiers \citep[e.g.][]{Richards-2011}. Misclassification, fuzzy boundaries between variable stars' classes, mis-labelled training sets, as well as, full processing of terabytes of data are current scientific challenges \citep[][]{Eyer-2006}.
 
The present paper is the first in a series of papers covering different aspects of variable star selection and classification. The first two articles are related to selection of variable stars using variability indices. In this paper, we discuss the selection of variable stars using correlation variability indices, while in the second of this series we will discuss non-correlation variability indices; Paper 3 will be about periodicity search methods; Paper 4 will be about the variable star classifier. In this work, we perform a comprehensive stellar variability analysis on time varying data.  In Sect.~\ref{wfcam}, we describe the data used to compare each index, using a pre-selected catalogue of known variable stars to test how well each index selects these and the efficiency of the selection measured by how few additional stars are selected by the same cutoff value. In Sect.~\ref{varind}, we present an overview of commonly used correlation variability indices and propose 5 new variability indices. Next, in Sect.~\ref{new_idea} we analyse the limits of correlated variability indices as well as proposing a false alarm probability for variability indices. We present our results and discussions in Sect.~\ref{res}. Finally, in Sect.~\ref{bestsel}, we draw our conclusions and discuss some future perspectives.

\section{Data} 
 
\subsection{WFCAMCAL database}\label{wfcam}

The public WFCAM Calibration \citep[WFCAMCAL - ][]{Hodgkin-2009,Cross-2009} is an unique programme that is well fitted to test the panchromatic variability indices and our assumptions. This programme contains panchromatic data for 58 different pointings distributed over the full range in right ascension and spread over declinations of $+59\fdg62$ and $-24\fdg73$. These were used to calibrate the UKIDSS surveys \citealt[][]{Lawrence-2007}. The pointing closest to the zenith  was chosen whenever a calibration field was observed. This was typically every hour early on in the UKIDSS observations and later every 2 hours, with some early nights having many additional observations (up to 40 in a night). During each visit the fields were usually observed with a sequence of filters, either  through $JHK$ or $ZYJHK$ filters within a few minutes. This lead to an irregular sampling with fields observed again roughly on a daily basis, although longer time gaps are common, and of course large seasonal gaps are also present in the data set.

The WFCAMCAL data are archived in the WFCAM Science Archive (WSA; \citealt[][]{Hambly-2008}). The data are processed by the Cambridge Astronomy Survey Unit (CASU) \cite{Irwin-2004} and the Wide Field Astronomy Unit (WFAU) in Edinburgh, and the latter produce the WSA. The design of the WSA, the details of the data curation procedures and the layout of the database are described in detail in \citealt[][]{Hambly-2008} and \citealt[][]{Cross-2009}. We use data from the WFCAMCAL08B release (observations upto the end of UKIRT semester 08B).

\subsection{The WFCAMCAL Variable Star Catalogue}\label{wvsc1}

\citealt[][]{Ferreira-Lopes-2015} performed a comprehensive stellar variability analysis of the WFCAMCAL database and presented the photometric data and characteristics of the identified variable stars as the WFCAM Variable Star Catalogue (WVSC1). The authors used standard data-mining methods and introduced new variability indices designed for multiband data with correlated sampling. To summarize, the authors performed a careful analysis using cutoff surfaces to obtain a preselection with $6651$ stars based on criteria established by numerical tests of the noise characteristics of the data. Next they combined four frequency analysis methods to search for the real frequencies in the LCs in each waveband and in the chromatic LC, i.e. comprised of the sum of all broadband filters. Finally, they obtained a ranked list of the best periods for each method and selected the very best period, which gave the minimum $\chi^{2}$ in order to cope with aliasing. Finally, the authors visually inspected all the phase diagrams of the $6651$ stars and recovered a catalogue containing $319$ stars in which $275$ are classified as periodic variable stars and $44$ objects  as suspected variables or apparently aperiodic variables. 

In this paper we analyse this same sample from \citealt[][]{Ferreira-Lopes-2015}. First, we selected all sources classified as a star or probable star having at least ten unflagged epochs in any of the five filters. This selection was performed from an initial database of $216,722$ stars. Next we test the efficiency of selection of variable stars using the variability indices presented in Sect.~\ref{varind}.

\section{Variability Indices}\label{varind}

Table~\ref{tvar01} summarises 12 variability indices of which 5 are new indices proposed in this work. The present work discusses the efficiency of selection of each one and discusses the best way to select variable stars using the current tool inventory. Sets of variability indices have been used, instead of one, to improve the selection process during the last few years \citep[e.g.][]{Shin-2009}. Indeed automatic classifiers are also using these parameters to facilitate the classification of variable stars \citep[e.g.][]{Richards-2011}. The variability indices are a fundamental tool to improving all processes of the time domain analysis.

Currently, the Welch-Setson indices \citep[e.g.][i.e. $I_{WS}$, $J_{WS}$,
$K_{WS}$ and $L_{WS}$ indices]{Welch-1993,Stetson-1996} are found to be
significantly more sensitive than the ``traditional'' $\chi^2$-test for single
variance, which uses the magnitude-rms scatter distribution of the data as a 
predictor \citep[e.g.][]{Pojmanski-2002}. The improvements proposed by 
\citealt[][]{Stetson-1996} on $I_{WS}$ \citep[][]{Welch-1993} and incorporated 
in the $J_{WS}$ index allow us to compare wavebands with different numbers of 
epochs on an equal basis. The author uses the Bessel correction
($\sqrt{\frac{n}{n-1}}$) to reduce the bias related with the sample size despite
the index being the square of the correlation not the mean variance. The
$I_{WS}$ index was modified, to quantify panchromatic flux correlations, to
form new variability indices ($I_{\rm pfc}^{(s)'}$) by 
\citealt[][]{Ferreira-Lopes-2015}. These were the first variability indices
developed to analyse panchromatic surveys. Moreover the authors proposed a new  
set of flux independent variability indices ($I_{\rm fi}^{(s)}$). 

The statistical period search based in the analysis of variance 
\citep[ANOVA,][]{Schwarzenberg-Czerny-1996} has been used to select
non-stochastic variations. Nevertheless, this method is limited to
identification of periodic variations and requires more running time once its  
significance level is determined on phase diagrams for each frequency test. 
Using variability indices we can discriminate non-stochastic variations 
independently from their nature and reduce the running time. The main goal of
this work is to determine the best way to select variable stars without  computing the variability periods. In the follow subsection we summarize the  $I_{\rm pfc}^{(s)'}$ and $I_{\rm fi}^{(s)}$ variability indices as well as improvements on these indices using a new approach.

\subsection{The $I_{\rm pfc}^{(s)'}$ and $I_{\rm fi}^{(s)}$ panchromatic variability indices}\label{ourind}

The current tool inventory was added to by \citealt[][]{Ferreira-Lopes-2015} with a new set of variability indices to separate LCs that are dominated by correlated variations from those that are noise-dominated ones. The authors introduced a new set of variability indices designed for multi-band data with correlated sampling that included one index that is highly insensitive to the presence of outliers in the time-series data. First, the authors extended $I_{\rm WS}$ to create the $I_{pfc}$ index defined as,

\begin{equation}
      I_{\rm pfc}^{(s)'} = \! \frac{1}{n_{s}}\sum_{i=1}^{n}\left [  \sum_{j_1=1}^{m-(s-1)}\!\!\!\cdots \left ( \sum_{j_s=j_{(s-1)}+1}^{m} \!\!\! \Lambda_{ij_1\cdots j_{s}}^{(s)}  \sqrt[s]{\left| \Gamma u_{ij_1} \cdots \Gamma u_{ij_s}  \right|} \right ) \right ],
   \label{eq_pfcgen}     
\end{equation}   

\noindent where $m$ is the number of filters, $s$ is the combination type (between two or more epochs), $n_{s}$ is the total number of correlations, and the $\Lambda^{(s)}$ correction factor is,

\begin{equation}
      \Lambda_{ij_1\cdots j_{s}}^{(s)} = \left\{ \begin{array}{ll}
      +1 & \qquad \mbox{if \,\, $\Gamma u_{ij_1} > 0, \,\, \cdots \,, \, \Gamma u_{ij_s} > 0$ };\\
      +1 & \qquad \mbox{if \,\, $\Gamma u_{ij_1} < 0, \,\, \cdots \,, \, \Gamma u_{ij_s} < 0$ };\\
      -1 & \qquad \mbox{otherwise}.\end{array} \right. 
   \label{eq_feps}    
\end{equation} 

\noindent and $\Gamma$ is given by,

\begin{equation}
    \Gamma u_{ij} =   \sqrt{ \frac{ n_{u_{j_{s}}} }{n_{u_{j_{s}}} - 1}} \times \left( \frac{u_{ij_{s}} - \bar{u}_{j_{s}}}{\sigma_{u_{ij_{s}}}} \right).
\label{eq_pkind}     
\end{equation} 

These indices allow us to compute correlations among $s$ epochs. As shown by the authors, increasing the number of correlated wave bands (s) makes the separation between correlated and uncorrelated variables more evident. Next the authors proposed a new index, $I_{fi}^{(s)}$, using Eqn~\ref{eq_feps} that is the sum of discrete values 1 or -1. This index is defined as

\begin{equation}
      2 \cdot I_{\rm fi}^{(s)} -1 = \frac{1}{N_{s}}\sum_{i=1}^{n}\left [  \sum_{j_1=1}^{m-(s-1)}\cdots \left ( \sum_{j_s=j_{(s-1)+1}}^m \Lambda_{ij_1\cdots j_{s}}^{(s)} \right ) \right ]\,,
   \label{eq_figen}     
\end{equation}   

\noindent where $0 \leq I_{\rm fi}^{(s)} \leq 1$, and where $\Lambda_{ij_1\cdots j_s}^{(s)}$ is defined in Eqn~(\ref{eq_feps}). Finally the authors propose a general expression to determine the probability of a random event leading to a positive $I_{\rm fi}^{(s)}$ index. In the case of statistically independent events, this is given by,

   \begin{equation}
     P_{s} = \frac{2}{s^{2}}. 
   \label{prob002}     
   \end{equation}

On the other hand the expected value of $I_{\rm pfc}^{(s)'}$ for a random distribution is about $0$. Meanwhile, the number of sources with negative values increase with $s$, because there is an increase in the number of possible combinations that give a negative correlation.

\subsection{Improvements on panchromatic and flux independent indices}\label{improv_ind}

The $I_{\rm pfc}^{(s)'}$ variability indices (see Eqn.~\ref{eq_pfcgen}) would work equally well if a set of observations are in the same bandpass, if we correlated groups of observations observed over a short interval. Therefore we need to modify the $I_{\rm pfc}^{(s)'}$ indices to make them still more robust against different numbers of observations in each group. Similarly, we propose a new panchromatic, flux independent, variability index ($K_{\rm fi}^{(s)}$) and additionally combine these indices to create two new indices. In order to provide an expression to be used in multi or single waveband data we 
propose the follow notation:

\begin{enumerate}

\item First, we compute the values of  $\delta_i$ give by
\begin{equation}
   \delta_i = \sqrt{\frac{n_{x}}{n_{x}-1}}\cdot \left( \frac{x_i-\bar{x}}{\sigma_{x,i}} \right).
\label{delta_stet}     
\end{equation}
  
\noindent where $n_{x}$ is the number of epochs of waveband $x$, $x_{i}$ are the flux measurements, $\bar{x}$ is the mean flux and $\sigma_{x,i}$ denotes the flux errors. This parameter is equal to that used by \citealt[][]{Stetson-1996} to improve $I_{WS}$ index and according to him the measurements of correlations using $\delta_i$ allow us to compare data from different wavebands with unequal numbers of observations on an equal basis.

\item Next, the $\delta_i$ values are computed for all measurements in each
waveband using the respective values of $n_{x}$. As a result we obtain a vector $\delta$ with $N$ measurements collected in any waveband. In addition, we save the observation time for each $\delta_i$ value.

\item We determine the time interval ($\Delta T$) for which measurements enclosed in this interval will be considered to be at virtually the same epoch. The chosen $\Delta T$ value comes from the arrangement of epochs and thus gives the minimum period (see Sect.~\ref{secdett}). The total number of boxes ($N_{box}$) is given by $T_{tot} / \Delta T$, where $T_{tot}$ is the  total time spam. The accuracy of the index must increase as $\Delta T$ decreases.

\item Next we compute the value of the variability index of order $s$ in the $k^{th}$ box;
   
\begin{equation}
    Q_{(s,k)} = \left\{ \begin{array}{ll}
              \sum_{j_1=1}^{n_k-(s-1)}\!\!\!\cdots \sum_{j_s=j_{(s-1)}+1}^{n_k}\left(  \Lambda_{j_1,\!\!\!\cdots,j_s}\!\!\! \cdot \sqrt[s]{ \left| \delta_{j_1} \!\!\!\cdots \delta_{j_s} \right|} \right)  & \quad \mbox{if $j_1 \neq j_s$} ;\\
             0 & \quad \mbox{if $n_{k} \le 1$.}  \end{array} \right.     
\label{eq_pkind}     
\end{equation}   

This equation performs all possible combinations without repetition among the $n_k$ values. Indeed the total number of combinations calculated is given by,

\begin{equation}
   N_{s} = \sum_{k=1}^{N_{box}} \frac{n_k!}{[s!(n_k-s)!]}.      
\label{eq_njind}     
\end{equation} 
    
\item Now, we can express the flux independent indices on a simple expression given by,

\begin{equation}
    K_{\rm fi}^{(s)} = \frac{N_{s}^{+}}{N_{s}}
   \label{eq_kind}     
\end{equation}   

\noindent where $N_{s}^{+}$ are the number of positive correlations according to Eqn.~\ref{eq_feps}. $K_{\rm fi}^{(s)}$ indices include measurements obtained in one filter in contrast to $I_{\rm fi}^{(s)}$ indices where measurements are obtained in different filters. 

\item Next, we can compute the $L_{\rm pfc}^{(s)}$ index given by,

\begin{equation}
  L_{pfc}^{(s)} =  \frac{1}{N_{s}}\sum_{k=1}^{N_{box}} Q_{(s,k)}.
\label{eq_lind}     
\end{equation} 

\noindent where it reduces to $J_{pfc}^{(s)}$ in the case when we only
have measurements in different filters. $L_{pfc}^{(s)}$ can be used to perform
comparisons on an equal basis between stars with different number of epochs as
well as using measurements obtained in one filter in contrast to $J_{\rm
pfc}^{(s)}$ and $K_{\rm fi}^{(s)}$ indices.

\item An alternative way to compute the characteristic value of
correlation is computing the median of correlations, given by

\begin{equation}
  M_{pfc}^{(s)} =  med\left[ Q_{(s)} \right],
\label{eq_mind}     
\end{equation} 

where $Q_{(s)}$ encloses all $ Q_{(s,k)}$ correlations. The median value may provide a more robust value than the mean in the presence of outliers.

\item Finally, we can use the $K_{\rm fi}^{(s)}$ in a correction factor related to instrument properties and outliers. Such a factor can be defined as,

\begin{equation}
     F^{(s)} = \left\{ \begin{array}{ll}
             2 \cdot \left( K_{\rm fi}^{(s)} - P_{s} \right)   & \quad \mbox{if $K_{\rm fi}^{(s)} \geq P_{s}$} ;\\
             0 & \quad \mbox{otherwise}  \end{array} \right. 
   \label{eq_migen}     
\end{equation}   

\noindent where $P_{s}$ is the expected value of pure noise for $K_{\rm
fi}^{(s)}$. $F^{(s)}$ ranges from $0$ to $2\times(1-2/s^2)$ providing an
increase of its weight with s values. For instance, the maximum value
of $F^{(s)}$ is $1$ for $s = 2$ and $\sim 1.6$ for $s = 3$. $F^{(s)}$ is more
efficient than $K_{\rm fi}^{(s)}$ because this increases the difference between
values of correlated and uncorrelated data and we concentrate the pure noise
values about zero. $F^{(s)}$ is used for provide a new set of indices given by
$FL^{(s)}$ and $FM^{(s)}$ (see Table~\ref{tvar01}).

\end{enumerate}

The variability indices proposed above are determined using properties 
related with the magnitude and signal correlation values. Stellar variability 
searches based on such indices follow general assumptions: (i) intrinsic
stellar variability can be typically identifiable from analysis of correlation 
measures observed in multiple or single wavebands; (ii) there are a minimum
number of correlations required to discriminate stochastic and non-stochastic
variations (see Sect.~\ref{numme}); (iii) the interval between any 2 
observations ($\tau$) used to compute the correlations must be sufficiently 
phase-locked (see Sect.~\ref{secdett}); (iv) non-intrinsic variations will be 
typically stochastic. Indeed measurements due the systematics of instrumental 
and atmospheric origin, or due to possible data reduction anomalies, displaying 
correlated properties may decrease the confidence level of variability indices. 
Such measurements are mainly related  with temporal saturation of bright objects
as well as systematic variations in the sky noise for faint stars.

\begin{figure}[htb]
  \includegraphics[width=0.45\textwidth,height=0.3\textwidth]{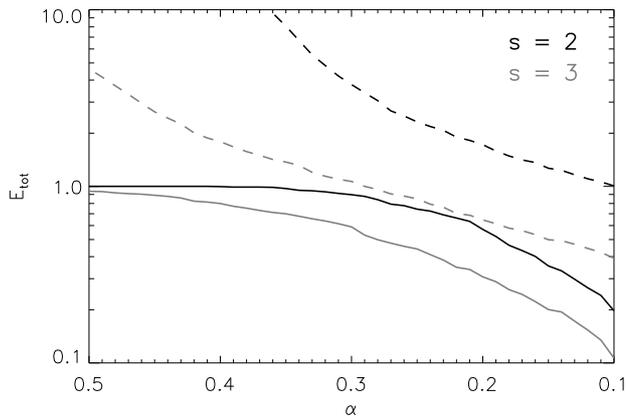}
  \caption{Efficiency metric ($E_{tot}$: the ratio of total number of sources 
  in the selection to good known variables in WVSC1) using Eqn.~\ref{corrupmes}
  for $s  = 2$ (black lines) and for $s = 3$ (grey lines). Solid lines mark 
  $E_{WVSC1}$, the fraction of the good variables in the selection, while the
  dashed lines mark $E_{tot}$. A good choice of $\alpha$ returns a high fraction
  of good variables $E_{WVSC1}$ for a low value of the efficiency metric $E_{tot}$.}
 \label{fignumsel}
\end{figure}

\section{Detection limits of correlated variability indices}\label{new_idea}

The number of measurements and how many measurements are `close' - i.e. within a
time span much shorter than the period of any variability - are fundamental 
information necessary to set better variability indices. The number of
measurements will determine how stringent the cutoff values must be while the 
number of close measurements will set the most appropriate variability index. To
determine which measurements are close or not we need to determine a $\Delta T$ 
value such that it is a compromise between the number of correlated measurements
and the minimum period that we are searching for. For instance, variability
indices computed in boxes of $\Delta T$, greater than the period, will return 
values closer to those expected for noise. Lower values of $\Delta T$ lead to 
higher accuracy for variability indices that use correlation measurements.

Moreover, variability indices can be used in all processes of the time-series analysis such as discussed in previous sections. Therefore we must find new ways that allow us to increase the precision, reliability of these indices and their connexions with the different types of variability. In Sect.~\ref{varind} we described the current tool inventory and we proposed new variability indices with new correction factors to reduce bias. In the present section we propose new ways to increase the precision of these variability indices as well as how to evaluate their reliability.

\subsection{Number of correlated measurements}\label{numme}

The minimum number of measurements that are enough for the use of a variability index will be determined by the capability of separating variable and non-variable stars. The statistical properties like mean, standard deviation, skewness and kurtosis will be strongly dependent on the number of measurements. On the other hand, we need contemporary (close) measurements to use correlated variability indices. These features may change for each variability index.

Consider two cases: one with $N_{s}$ correlated data points and the other with $N_{s}$ of pure noise for $s = 2$. In the case of pure noise, the number of positive and negative correlation must be the same $(N_{s}^{+} = N_{2}/2)$, while for correlated data $(N_{s}^{+} = N_{2})$. Using $K_{\rm fi}^{(s)}$ indices we can determine the minimum number of correlations necessary to separate a purely correlated signal from pure noise assuming that there is an uncertainty in the sign of some correlations. We assume that this uncertainty (or fluctuation) on the number of positive correlations given by $n_f$ provides an increase in the variability index of pure noise and a decrease otherwise. So the minimum separation between correlated and uncorrelated data is given by,
 
\begin{equation}
   \Delta K_{\rm fi}^{(2)} = \left(\frac{N_{2} - n_f}{N_{2}}\right)- \left(\frac{\frac{N_{2}}{2} + n_f}{N_{2}}\right) > 0 \quad  \Rightarrow  \quad   N_{2}^{min} = 4\cdot n_f 
   \label{eqnp}     
\end{equation}    

\noindent where $n_f$ is an integer with values less than $N_{2}/2$. The minimum number of correlations must be at least 5 according to this relation, given a single error. The general expression is given by,

\begin{equation}
   \Delta K_{\rm fi}^{(s)} = \left(\frac{N_{s} - n_f}{N_{s}}\right)- \left(\frac{P_{s}N_{s} + n_f}{N_{s}}\right) > 0 \quad  \Rightarrow  \quad   N_{s}^{min} = \frac{2n_f}{1-P_{s}},
   \label{eqgen}     
\end{equation}    

\noindent where the minimum value ($\Delta K_{\rm fi}^{(s)}$) to validate this relation is given by,

\begin{figure}[htb]
  \includegraphics[width=0.45\textwidth,height=0.3\textwidth]{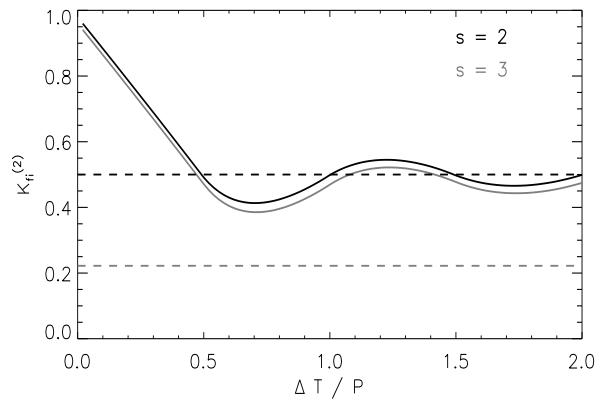}
  \caption{$K_{fi}^{(s)}$ versus $\Delta T / P$ for $s = 2$ (black lines) and for $s = 3$ (grey lines). The dashed lines mark the expected values for  random variation.}
 \label{figdett}
\end{figure}

\begin{equation}
   \Delta K_{\rm fi}^{(s)} = 1 - P_{s} - \frac{2n_f}{N_{s}}.
   \label{deltaLgen}     
\end{equation}    

\noindent where $n_f/N_{s}$ is the fractional fluctuation of positive correlated measures ($f_{\rm fluc}$). This equation explains analytically the increase in precision as well as the detachment between correlated and non-correlated data with increasing $s$ (see Figure 8 of \citealt[][]{Ferreira-Lopes-2015}). $\Delta K_{\rm 
fi}^{(s)}\rightarrow 1$ with the increasing $s$, so correlated data becomes more easily separable from pure noise with increasing $s$.

\subsection{False Alarm Probability on variability indices}\label{fap}

The statistical significance of a value is associated with the False Alarm Probability (FAP). FAP is the probability that the observed value was caused by random fluctations. The smaller the FAP then the larger will be the statistical significance of this measurement and the tolerance usually adopted is about $1\%$. The determination of FAP in period searches is hindered due to non-Gaussian distributions, observations scattered irregularly over a long time span, the unclear meaning of the number of independent frequencies and the manner in how these should be taken into account \citep[][]{Suveges-2014}.

Additionally, significance values may depend on the function employed to make the periodicity search. Therefore, in some cases, it indicates the use of different techniques on different types of variable stars \citep[][]{Templeton-2004}. For instance, the periodicity search methods based on Fourier series will be less sensitive to non-sinusoidal and aperiodic signals. Recent work based on the analyse of variance \citep[][]{Schwarzenberg-Czerny-1996} and multiharmonic periodograms \citep[][]{Baluev-2009,Baluev-2013} are allowing us to assess more complex signals more easily. This process may be facilitated if we can first determine whether the time series has reliable variability or not.

We can consider the significance of the variability indices by comparing the null hypothesis $H_0$ of the observed time series (purely noise) against the alternative $H_1$ stating that there is no correlated signal in it. One way to evaluate statistical fluctuations on variability indices is to generate a large number of test time-series sequences by shuffling the times (``bootstrapping''). Following this approach, we are able to keep part of the correlated nature of the  noise intrinsic to the data, as opposed to numerical tests based on pure Gaussian noise \citep[][]{Ferreira-Lopes-2015}. However, we need many iterations to provide accurate values for a FAP, i.e. we need  to compute the variability indices n times. For instance, at least 100 iterations are necessary to get a FAP of $0.01$ and this implies a longer running time.

\begin{figure}[htb]
  \includegraphics[width=0.45\textwidth,height=0.3\textwidth]{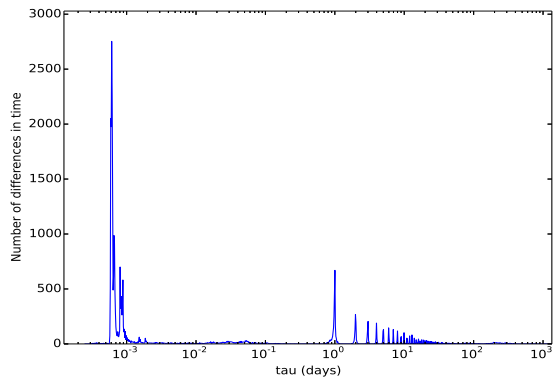}
  \caption{Histogram of the interval between observations for the WFCAMCAL08B data. tau is the interval between any 2 observations regardless of filter and these are binned logarithmically. The peaks at $\sim10^{-3}$ days are intervals during a $ZYJHK$ sequence and the peaks at $\sim1$ day and multiples of $1$day are repeat observations on subsequent nights.}
 \label{figCALcad}
\end{figure}

On the other hand an analytical expression of FAP for variability indices may depend on the deviation from the mean which will vary according to the survey analysed since it must depend on detector efficiency, number of measurements, magnitude, etc. However, $K_{\rm fi}^{(s)}$ has a weak dependence on the properties of the survey in which Eqn.~\ref{deltaLgen} provides a value above which correlated sources may be distinguished from noise. The only term to be determined is the fractional fluctuation in positive correlated measurements ($f_{\rm fluc}=n_f/N_{s}$). From our results, we propose the following empirical equation,

\begin{equation}
    f_{\rm fluc} = \frac{n_{f}}{N_{s}} = \alpha -  \sqrt{\frac{\beta(\alpha)}{N_{s}}},
   \label{corrupmes}     
\end{equation} 

\noindent i.e. a constant ($\alpha > 0$) plus a term related to the number of correlations. The second term in this equation decays quickly to zero as $N_{s}$ increases and it provides a strong cutoff values on data with few epochs. $\beta < \alpha^{2} N_{s}^{Min}$ since the $f_{\rm fluc}$ for $K_{\rm fi}^{(s)}$ must be greater than 0 for any number of correlations. Large values of $\alpha$ give a more complete selection while smaller values result in a more reliable sample.
Figure~\ref{fignumsel} shows the number of sources selected using Eqn.~\ref{corrupmes} for $s = 2$ (black lines) and $s = 3$ (grey lines) as a function of $\alpha$ values. Solid lines mark the number of sources of WVSC1 stars, while the dashed lines mark the efficiency of selection (ratio of total number of sources to number of known variables). The total number of sources is normalization by the total number of WVSC1 stars (319) to give an efficiency metric ($E_{tot}$). Table~\ref{cutval} shows the number of total sources selected ($E_{tot}$) and the fraction of WVSC1 stars ($E_{WVSC1}$) for some $\alpha$ values. For instance, to select about $90\%$ of WVSC1 stars we need a sub-sample of $3.77\times 319$ stars using $\alpha = 0.30$ for $s = 2$. On the other hand to select about $92\%$ of WVSC1 stars we need a sub-sample of $3.71\times 319$ stars using $\alpha = 0.48$ for $s = 3$.

\subsection{$\Delta T$ estimate and correlated observations}\label{secdett}

The variation of variability indices with $\Delta T$ will depend on many factors such as: the variability period ($P$), the shape of the light curve, the signal-to-noise, and outliers. In order to estimate the influence of $\Delta T$ we simulate a pure sinusoidal variation with a period (P). Next, we compute the $K_{fi}^{(s)}$ indices and see how changing $\Delta T$ as a function of $P$ affects how well we can separate a sinusoidal signal from random noise.

Fig.~\ref{figdett} shows the $K_{\rm fi}^{(s)}$ indices as a function of $\Delta T/P$. $K_{fi}^{(s)}$ decreases quickly to the expected values for random variations when $s = 2$, while, when $s = 3$, $K_{\rm fi}^{(s)}$ remains higher than the expected noise value for all $\Delta T/P$. This result helps us to  understand and use $\Delta\,T$ values. For instance, if $\Delta\,T<0.1P$, we will get large values for $K_{\rm fi}^{s}$, clearly separated from noise and thus detect variability more easily.

\begin{table}[htbp]
 \caption[]{Efficiency metric for some $\alpha$ values.}\label{cutval}
\centering    
\begin{tabular}{c c c | c c c }        
\hline\hline  
\multicolumn{3}{c |}{ s = 2 }  & \multicolumn{3}{c}{ s = 3 } \\
\hline 
$\alpha$ & $E_{WVSC1}$ & $E_{tot}$ & $\alpha$ & $E_{WVSC1}$ & $E_{tot}$\\    %

0.20 & 0.57 & 1.72 & 0.20 & 0.31 & 0.65 \\
0.22 & 0.66 & 1.90 & 0.24 & 0.41 & 0.79 \\
0.24 & 0.72 & 2.18 & 0.28 & 0.50 & 0.96 \\
0.26 & 0.78 & 2.50 & 0.32 & 0.64 & 1.14 \\
0.28 & 0.84 & 3.06 & 0.36 & 0.71 & 1.42 \\
0.30 & 0.90 & 3.77 & 0.40 & 0.80 & 1.79 \\
0.32 & 0.93 & 4.83 & 0.44 & 0.88 & 2.44 \\
0.34 & 0.95 & 6.70 & 0.48 & 0.92 & 3.71 \\
0.36 & 0.99 & 9.58 & 0.52 & 0.95 & 5.86 \\
0.38 & 0.99 & 13.96 & 0.56 & 0.98 & 9.87 \\

\hline                                   
\end{tabular}
\end{table}

$\Delta\,T$ will be determined predominantly by the cadence of the data. For the
WFCAMCAL data, a sequence of 3 to 5 filters were observed over a period of 
$\sim0.005$ day, and then each pointing reobserved roughly 1 day later, with
longer intervals because of weather or seasonal limits on the observations, see 
Sect~\ref{wfcam}. This is displayed in Fig~\ref{figCALcad}, which shows the 
histogram of time between subsequent observations. There is a strong peak 
$\tau\sim10^{-3}$day which corresponds to a $ZYJHK$ sequence with duration of 
$\sim0.005$ day and a second peak at $\sim1$ day, and a variety of smaller peaks
at other durations, often a few days apart (bad weather, non-photometric nights)
and  a few small ones at long durations of tens or hundreds of days (field not
observed because it was too close to the Sun) or on timescales of 0.01 to 0.1
days: days when many calibration fields were taken for test purposes. The best 
choice of $\Delta T$ should be the minimum duration that encloses the correlated
observations, upto the end of the last integration. $\Delta\,T = 0.01$day is a
sensible choice as it is slightly wider than the typical box size so no observation is missed and allows us to look for variables with $P\geq0.1$day, although the main sampling  peak at $\sim1$ day may be expected to limit us to $P > 0.5$ days from the Nyquist frequency. Since the sampling is not rigidly at 1 day intervals shorter periods are possible. Having correlated sampling at more than 20 times the frequency of the main sampling rate will avoid additional constraints being applied to the period range.

However, if we have equally spaced data, we are severely restricted. If we have $s=2$ correlations and a spacing of $\tau$, then $\Delta\,T\geq2\tau$ and $P\geq20\tau$, if $s=2$. Given that at least 2 full periods are required for a confident identification of periodic behaviour, at least 40 observations would be required to constrain a very narrow range of periods. Thus, correlation indices become very inefficient for equally spaced data. Some deep extragalactic surveys have observations designed to maximise depth, so the observations are taken when seeing and sky levels are at the best, so the observation structure can be pseudo-correlated, but not on fixed time scales. In Paper 2 of this series we will discuss indices which work better with uncorrelated observations.

A correlated data set may be expected to have at least half of the $\tau$ values in a peak or small set of peaks (if several filters with slightly different exposure times) at around the correlation frequency and then the main sampling peaks at $\tau_{\rm samp}>20\tau_{\rm cor}$.

When we consider VISTA surveys, e.g. the VVV \citep{Minniti-2010}, the data are observed as pawprints, which then get co-added into tiles \citealt[][]{Cross-2012}, so there are always repeat observations on a short time-scale compared to repeat epochs. These observations are also observed almost contemporaneously, with some tiling patterns jumping between different the same jitter on different pawprints before moving  onto the next jitter\footnote{http://casu.ast.cam.ac.uk/surveys-projects/vista/technical/tiles}, so the time between pawprints will usually be shorter than the integration time of the pawprint, so these are ideal for correlated indices applied to the pawprints. 

Gaia \citep{Bailer-Jones-2013} is another mission where correlated indices will be extremely valuable. The main astrometric instrument observes stars as they transit across 9 strips of detectors with $\tau\sim5$s. Stars are then reobserved by a second field of view $2$h later or on the next revolution $6$h later or on longer time scales due to the orbit and precession of the spacecraft.

\begin{figure}[htb]
  \includegraphics[width=0.45\textwidth,height=0.4\textwidth]{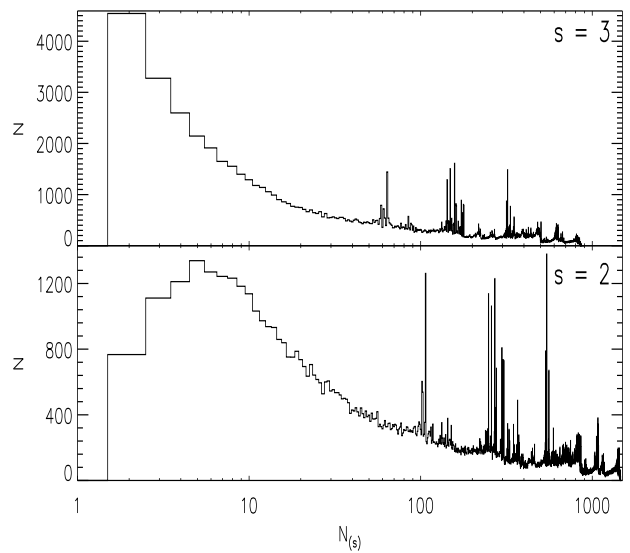}
 \caption{Histograms of the number of correlations $N_{s}$ for $s = 2$ and $s = 3$ using  bins of width 1.}
 \label{numcor}
\end{figure}

\begin{figure*}[htb]
 \includegraphics[width=0.9\textwidth,height=1.1\textwidth]{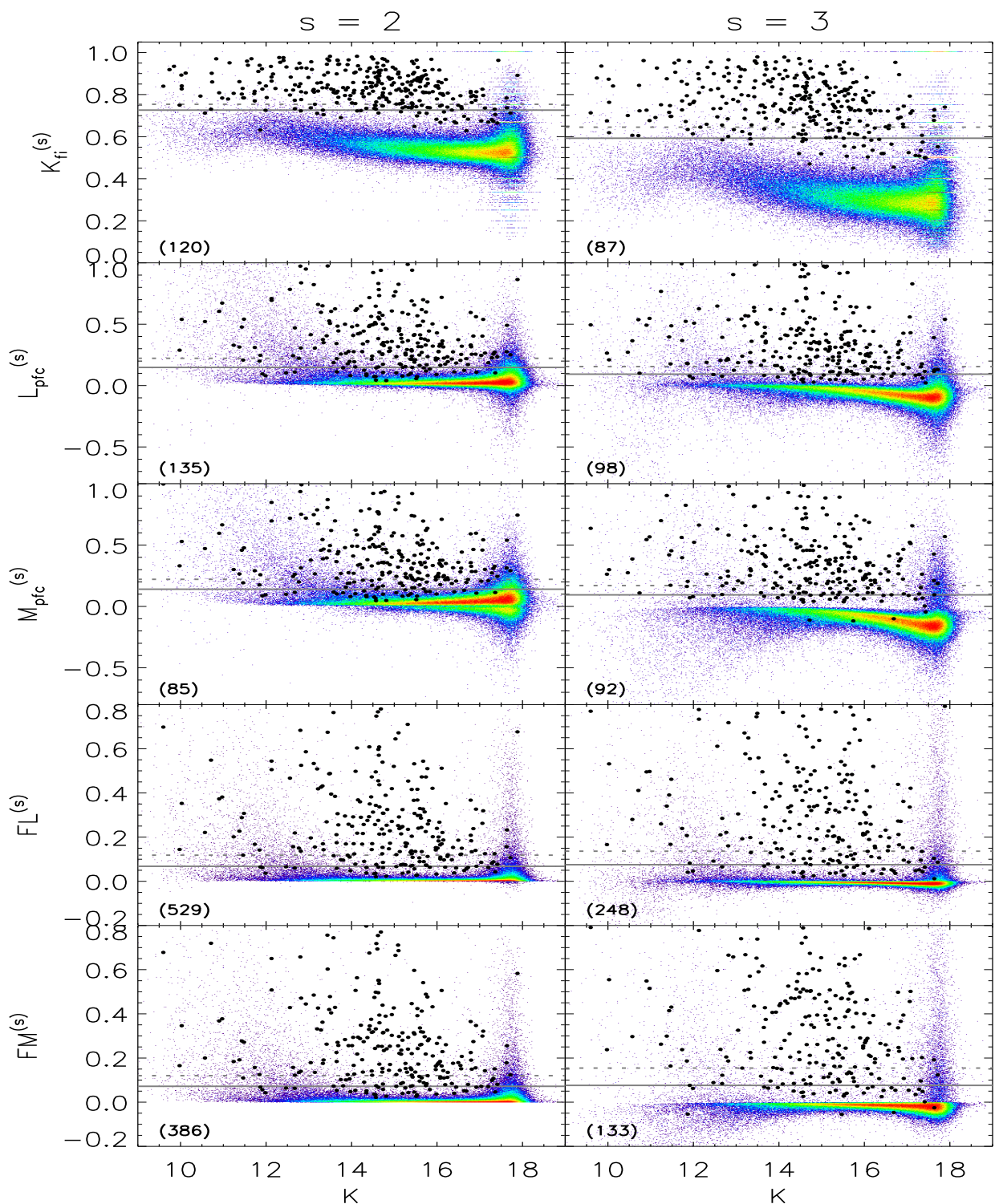}
 \caption{Correlation variability indices versus K magnitude (left diagram) and 
 versus number of correlations (right diagram) in each panel. The maximum number
 of sources per pixel is displayed in brackets in each panel. The black circles
 mark the WVSC1 sources and the solid and dashed lines marks the value which 
 encloses $90\%$ and $80\%$ of them.}
 \label{figindcor}
\end{figure*}

\section{Data analysis}\label{data}

\subsection{Broad selection and Bias}\label{mbias}

From Sects.~\ref{numme}, \ref{fap}, and \ref{secdett} we can determine
the main constraints on variability analysis. We consider that there is at least one incorrect correlation measurement $n_{f}=1$ in each LC therefore we limit our analyses to sources with more than four correlation measurements, according to Eqn~\ref{eqgen}. This is the minimum number of correlation measurements adopted in our analysis. Moreover, we adopted $\Delta\,T = 0.01$ days, based on the duration of the $ZYJHK$ sequences. By following these constraints, we are considering all LCs that can possibly discriminate a correlated signal from noise with periods of at least greater than $0.1$ days (see Sect.~\ref{secdett}) We revisited the WFCAMCAL data instead of testing these variability indices using simple sinusoidal light-curves, since this gives a more realistic test with correlated observations, real noise values, and a range of variable types.

Next, we compute the $K_{\rm fi}$, $L_{\rm pfc}$, $M_{\rm pfc}^{(s)}$, 
$FL^{(s)}$, and $FM^{(s)}$ variability indices using a
multi-waveband approach, as discussed in Sect.~\ref{improv_ind} on the data 
described in Sect.~\ref{wfcam}. Fig.~\ref{numcor} shows the histogram of the
number of correlation $N_{s}$ ranging from 1 to 1467 for $s = 2$, and from $1$ 
to $863$ for $s=3$. These numbers are different because the $ZYJHK$ measurements
are obtained within a few minutes of each other but not necessarily in all
filters. Additionally, the number of correlation measurements decreases quickly 
for very faint objects around the detection threshold. The total baseline varies
from a few months up to three years and the cadence in a single passband can be 
considered to be quasi-stochastic with rather irregular gaps \citep[see][for a
better discussion]{Hodgkin-2009,Cross-2009,Ferreira-Lopes-2015}.

\subsection{Searching for periodic variations}\label{secperiod}

To search for the best period a Lomb-Scargle periodogram 
\citep[][]{Lomb-1976,Scargle-1982} was computed for each LC. We set the 
low-frequency limit ($f_0$) for each periodogram to be $f_0 = 2/T_{\rm 
tot}$days$^{-1}$, where $T_{\rm tot}$ is the total time spanned by the LC. The 
high-frequency limit was fixed to $f_{N} = \frac{1}{\Delta\,T} =
10$~days$^{-1}$, and the periodogram size was scaled to $10^{5}$ elements. 
Initially, we compute the Lomb-Scargle periodogram independently for each broadband filter ($Y$, $Z$, $J$, $H$, and $K$) as well as for the chromatic light curve (as described in \citealt[][]{Ferreira-Lopes-2015}). The use of all broadband filters allows us to find variability periods in the cases where the photometry is high quality in some filters, but not others: e.g. in some filters the object may be saturated (sometimes leading to a non-detection at the correct location), or may be very faint, close to the detection limit or even too faint to be detected.

For each broadband filter as well as for the chromatic light curve we retain the $10$ periods corresponding to the highest peaks. Next these periods were refined following \citet{De-Medeiros-2013}, namely, by maximizing the ratio of the variability amplitudes to the minimum dispersion in the phase diagram given by \citet[][]{Dworetsky-1983}. Finally, in order to select the very best period, we use the $\chi^{2}$ test, in the same way as described in \citealt[][]{Ferreira-Lopes-2015}.

\begin{figure}[htb]
  \includegraphics[width=0.48\textwidth,height=0.5\textwidth]{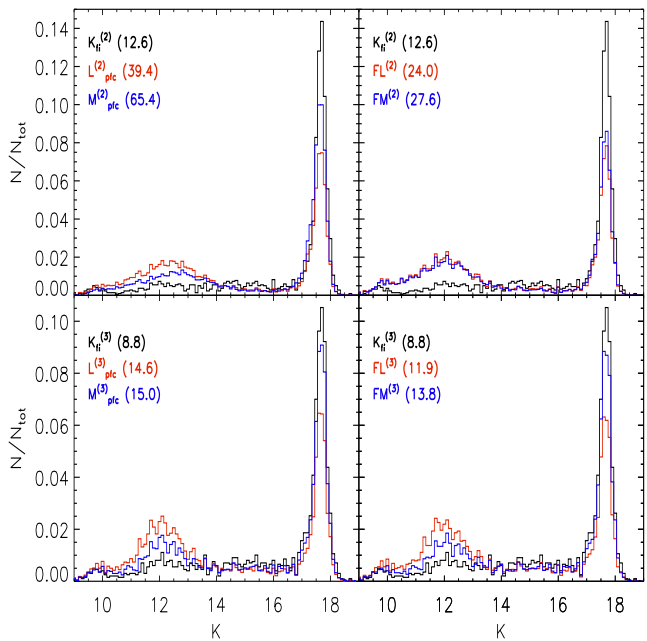}
 \caption{Histograms of the sources selected with a constant cutoff value for  $K_{\rm fi}^{(s)}$ (black line), $L^{(s)}_{pfc}$ and $FL^{(s)}$ (red line), and $M^{(s)}_{pfc}$ and $FM^{(s)}$ (blue line) indices normalized for the total number of sources  selected in each one of them. $E_{tot}$ values for each index is displayed in parentheses. The upper panel shows the histograms for $s = 2$ while the lower panel shows them for $s = 3$. The cutoff values were determined considering a value that includes $90\%$ of WVSC1 stars (see Table~\ref{numsel}).}

  \label{histsel}
\end{figure}

\section{Results and Discussions}\label{res}

We analyse the efficiency of variability indices for selecting variable stars in the WFCAMCAL database. We evaluate responses of these indices as a function of magnitude and the number of correlations. This study allows us to trace important remarks about the most efficient way to select variable stars. The most efficient index is the one that encloses the majority of the WVSC1 stars with the fewest stars which do not belong to the WVSC1 catalogue and are mostly misclassifications. We compute the variability indices as described in Sect.~\ref{improv_ind}. We detail our results in an analysis of correlation variability indices that were computed using a panchromatic approach such as described in Sect.~\ref{improv_ind}. Below we present our results using stars from the WVSC1 catalogue as a comparison.

\begin{figure*}[htb]

\includegraphics[width=0.5\textwidth,height=0.4\textwidth]{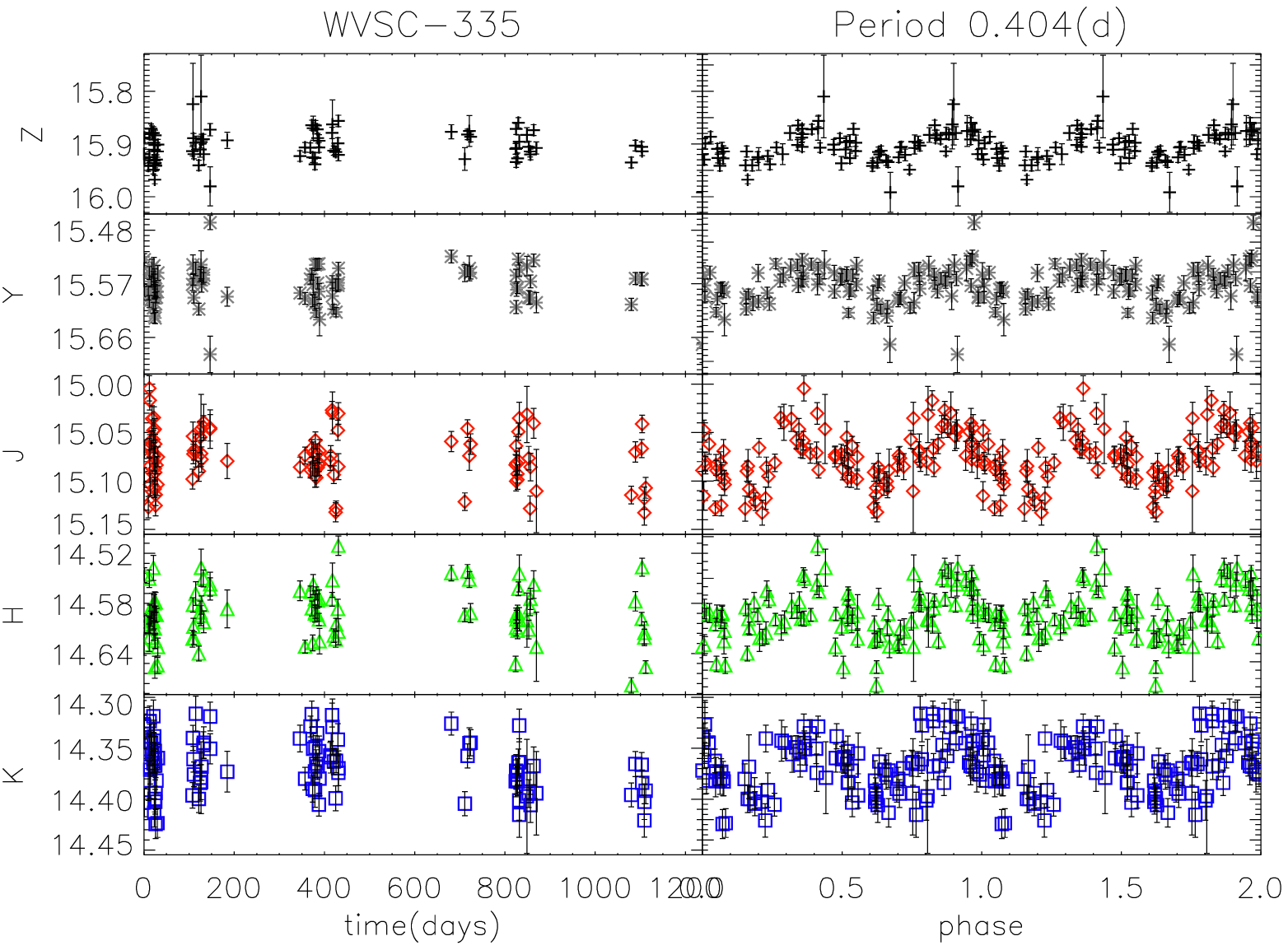}
\includegraphics[width=0.5\textwidth,height=0.4\textwidth]{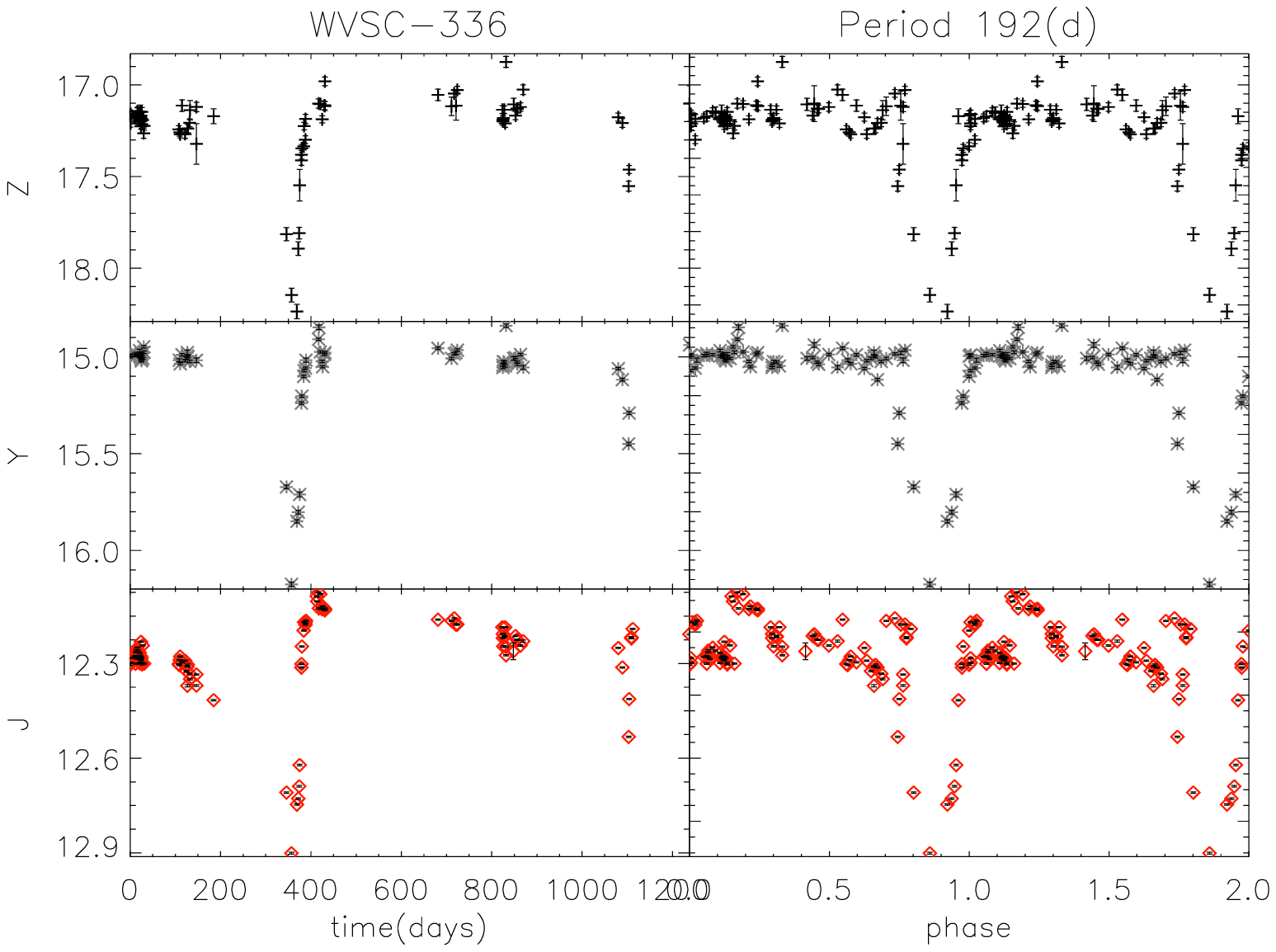}

\includegraphics[width=0.5\textwidth,height=0.4\textwidth]{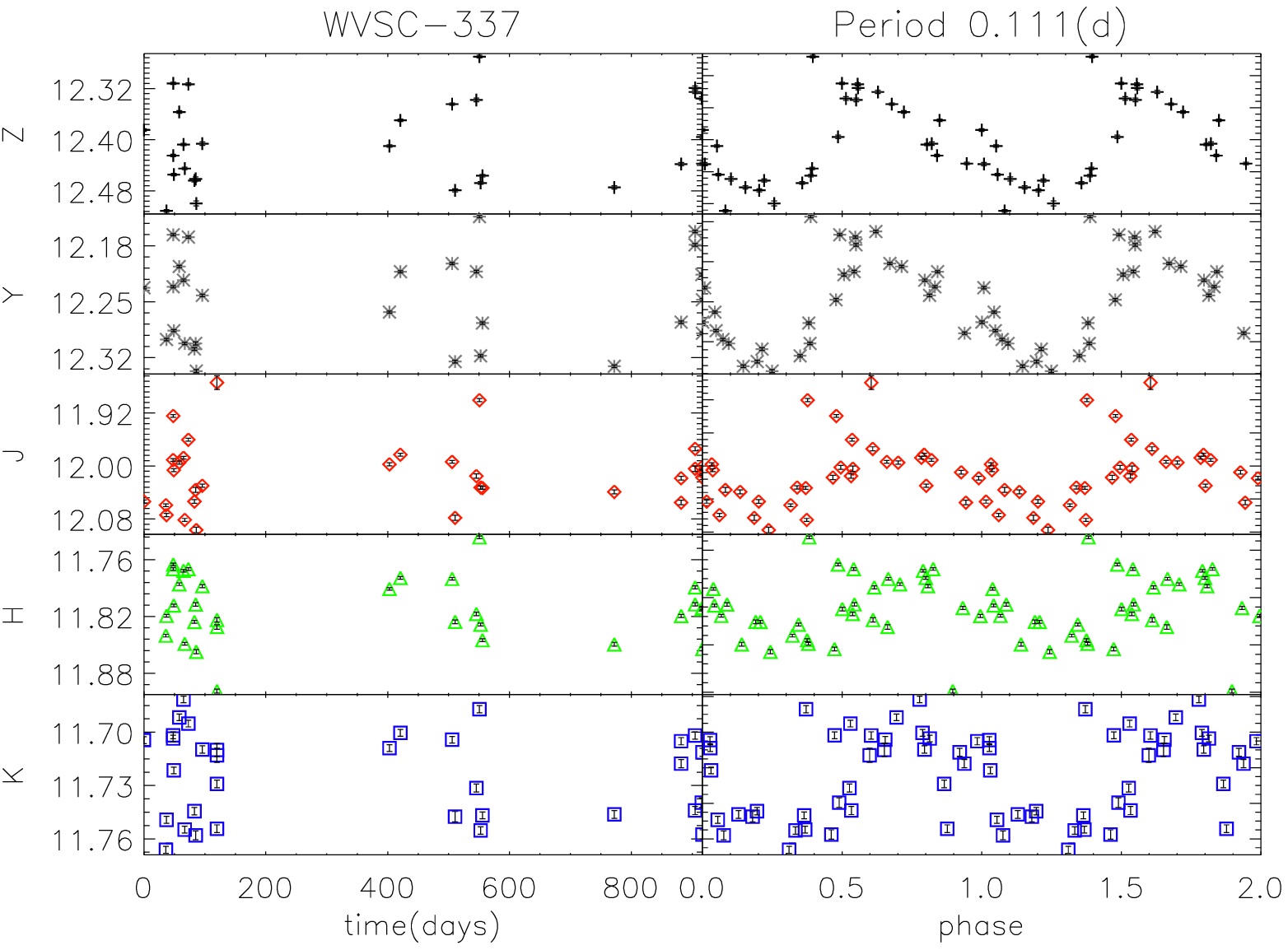}
\includegraphics[width=0.5\textwidth,height=0.4\textwidth]{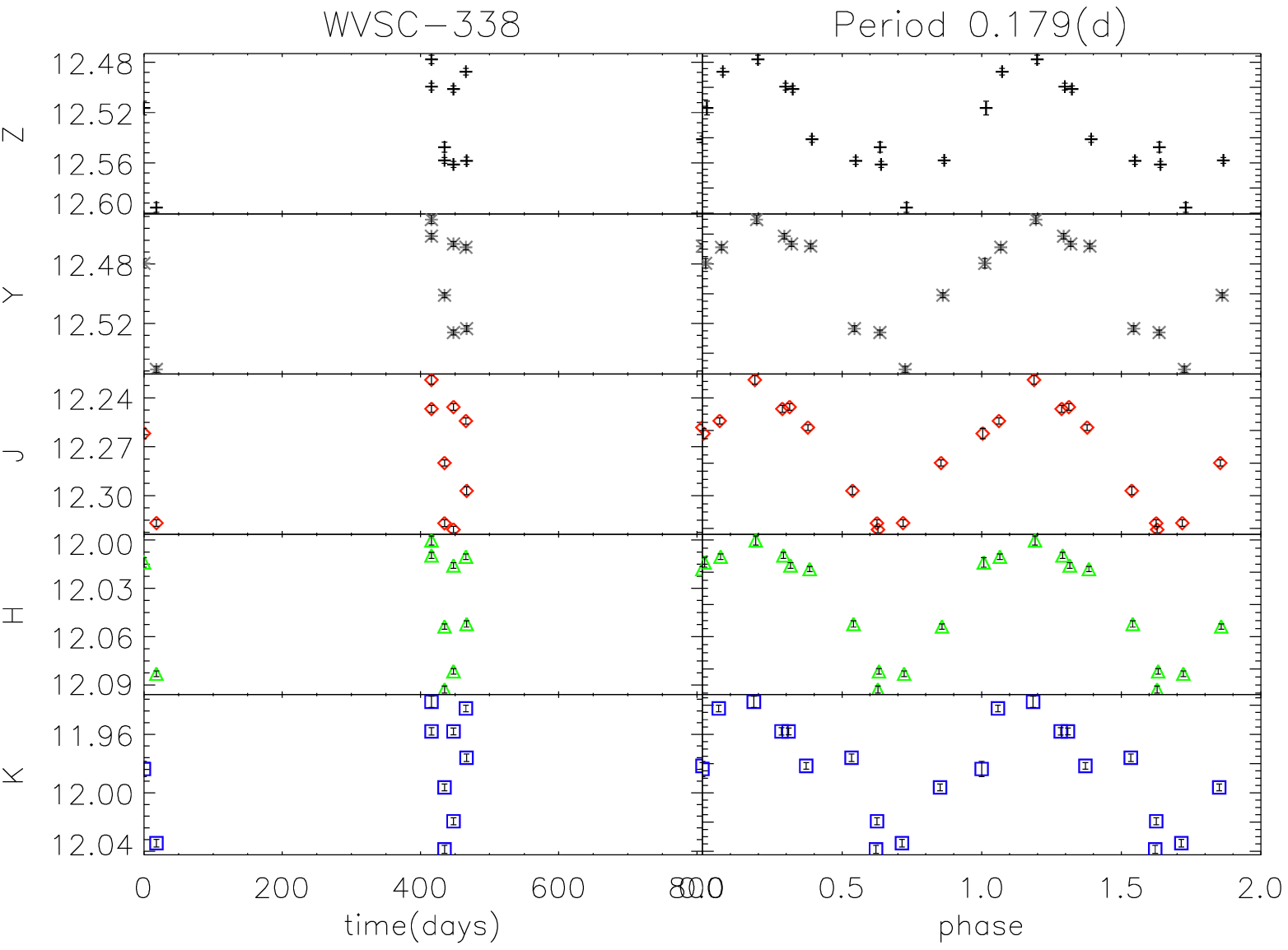}

 \caption{LCs and phase diagrams of C1 catalog. The identifiers and periods are displayed above each panel.}
 \label{figlcvar}
\end{figure*} 

\subsection{Efficiency of variability indices}\label{rescor}

Fig~\ref{figindcor} shows the distribution of $K_{\rm fi}$, $L_{\rm pfc}$, 
$M_{\rm pfc}^{(s)}$, $FL^{(s)}$, and $FM^{(s)}$ variability indices as a
function of magnitude and the number of correlations $N_{s}$, for $s=2$ and
$s=3$. The solid and dashed lines are set to the values that must be
adopted if we want to select $90\%$  and  $80\%$ of the final WVSC1 catalogue, 
respectively. Fig~\ref{histsel} shows the histogram of these indices as a
function of magnitude for stars selected using the cutoff value that includes 
$90\%$ of the WVSC1 catalogue.] 
  
\begin{itemize}

\item $K_{\rm fi}^{(s)}$ presents a clear separation of WVSC1 stars from the other stars for $K \leq 17$ mag. The lines that appear for $K \geq 17$ mag are due to the high number of sources with just a few epochs (typically $N_{s}<20$). This index produces discrete values and these are more evident for a small number of epochs. The right panel shows a higher dispersion for low numbers of $N_{s}$ as expected. Statistical fluctuations may provide high contamination in this region despite the number of correlations being above the minimum number 
that allows us to discriminate them, according to Eqn~\ref{eqgen}. $K_{fi}^{(3)}$ shows a similar behaviour although it displays a greater separation of WVSC1 stars from the other stars.

\item $L_{\rm pfc}^{(s)}$ is equivalent to two previous indices under
some constraints: it is equal $J_{\rm WS}$ for $s = 2$ and equivalent to $J_{\rm
pfc}^{(s)}$ when the correlations are obtained in different filters. $J_{WS}$
has been used in the selection criteria for several surveys 
\citep[e.g.][]{Christiansen-2008,McCommas-2009,Morales-Calderon-2009,
Bhatti-2010,Shappee-2011,Pasternacki-2011}.

$L_{\rm pfc}^{(s)}$ provides a higher selection efficiency than previous
indices and it performs combinations among measurements in $\Delta T$ intervals 
rather than across wavelengths. $L_{\rm pfc}^{(2)}$ indices present a clear
separation of WVSC1 stars from other stars. If we assume a constant value of
$L_{\rm pfc}^{(s)}$ as a selection criterion we observe that most of
the non-variable stars selected are faint stars. The separation between WVSC1
and other stars is clearer for $s=3$ than for $s=2$. Indeed, the number of
sources preselected to enclose $90\%$ of WVSC1 stars are about $30\%$
fewer for $s=3$ than $s=2$. Two main features that are expected with increasing
$s$ are observed: the increase in the number of stars with negative indices
values and a better discrimination between variable and non-variable stars.

\item $M_{\rm pfc}^{(s)}$ calculates the median of the
correlation values in contrast to the mean encapsulated by $L_{\rm pfc}^{(s)}$.
Both mean and median values are used to determine the central or typical value
in a statistical distribution. The weight of the outliers 
is reduced in the median compared to the mean. Outliers in photometric data are
commonly associated with variations in brighter stars non-linearity and
saturation. On the other hand, the increasing dominance of correlated noise 
is expected for faint stars as the errors become dominated by the sky noise, 
rather than photon statistics. Fig.~\ref{histsel} shows a underestimation of $L_{\rm pfc}^{(s)}$) and $M_{\rm
pfc}^{(s)}$ indices implying on increase of misclassification. Indeed, the 
number of stars is higher of $M_{\rm pfc}^{(s)}$ than $L_{\rm pfc}^{(s)}$ for
faint stars and against for brighter stars. Fig.~\ref{histsel}
shows an increase in the fractions of stars selected at both the bright and
faint magnitudes for $L_{\rm pfc}^{(s)}$) and $M_{\rm pfc}^{(s)}$ indices,
implying an increase in the misclassification rate. The misclassification rate
is higher for faint stars when using $M_{\rm pfc}^{(s)}$ than when using $L_{\rm
pfc}^{(s)}$, and vice-versa for bright stars. Therefore stars that match both
criteria should have a lower misclassification rate at both the bright and faint
ends, so agreement between these indices should be considered as a selection
criteria.

Using larger $s$ values gives less weight to outliers and partially-correlated noise, and this leads to a better estimations of the centre of the distribution for both the mean and median. Therefore, the efficiency of $L_{\rm pfc}^{(s)}$ and $M_{\rm pfc}^{(s)}$ will increase and they tend to have similar $E_{tot}$ values for different indices for higher $s$ especially if sources with small numbers of correlations are removed, as observed in Table~\ref{numsel} for $s = 3$ and $N_{s} > 20$. Meanwhile, the  $K_{\rm fi}^{(s)}$ are the best indices to perform a selection of variable stars, when we only consider higher $s$ values as well as only those  sources with $N_{s} > 20$ (see Table~\ref{numsel}).
 
\item $FL^{(s)}$) and $FM^{(s)}$ provide better efficiency values ($E_{tot}$) among those indices computed from correlation magnitudes (see Table~\ref{numsel}). The $F^{(s)}$ factor provides a concentration of non-variables with values around zero as well as a reduction  in the spread of bright sources. We observe a reduction of more than $300\%$  on the number of sources pre-selected by $L_{\rm pfc}^{(s)}$  and $M_{\rm pfc}^{(s)}$ indices for $s = 2$ and about $20\%$ for $s = 3$ (see Fig.~\ref{histsel}). The large reduction is not found for $s = 3$ because these indices become more accurate with increasing $s$ values and this therefore decreases the weight of $F^{(s)}$. On the other hand, only a slight decrease in $E_{tot}$ was observed when we use as correction factor $K_{\rm WS}/0.789$ (see Table~\ref{numsel}). Such factor is used to build the $L_{\rm WS}$ Stetson index \citep[][]{Stetson-1996} that can be expressed by $L_{\rm WS} \approx L^{(2)}_{\rm pfc}\times K_{\rm WS}/0.789$.

\end{itemize}

\begin{table}[htbp]

 \caption[]{Efficiency metric ($E_{tot}$) for variability indices analysed to select $90\%$ and $80\%$ of WVSC1 stars for $N^{(s)} > 4$ and $N^{(s)} > 20$, respectively.}\label{numsel}
 \centering    
 \begin{tabular}{c | c c | c c}        
 \hline\hline                 
  \multicolumn{1}{c|}{ } & \multicolumn{2}{c|}{ $N_{s} > 4$ } & \multicolumn{2}{c}{ $N_{s} > 20$ }    \\
  \hline    
  Index &  $E_{tot}(90\%)$  &  $E_{tot}(80\%)$  & $E_{tot}(90\%)$  &  $E_{tot}(80\%)$  \\    
 \hline                        

$K_{\rm fi}^{(2)}$ & 12.6 & 8.8 & 4.7 & 2.9 \\
$K_{\rm fi}^{(3)}$ & 8.9 & 5.2 & 3.0 & 1.7 \\
$L_{\rm pfc}^{(2)}$ & 40.1 & 21.5 & 27.3 & 14.7 \\
$L_{\rm pfc}^{(3)}$ & 14.7 & 8.8 & 8.6 & 5.5 \\
$M_{\rm pfc}^{(2)}$ & 65.5 & 29.7 & 48.4 & 20.0 \\
$M_{\rm pfc}^{(3)}$ & 15.0 & 9.9 & 7.4 & 4.9 \\
$FL^{(2)}$ & 24.0 & 13.2 & 14.6 & 7.9 \\
$FL^{(3)}$ & 12.0 & 7.4 & 6.7 & 4.4 \\
$FM^{(2)}$ & 27.6 & 15.1 & 16.7 & 8.9 \\
$FM^{(3)}$ & 13.8 & 8.5 & 6.5 & 4.0 \\
$L_{\rm WS}$ & 38.4 & 20.1 & 25.6 & 13.2 \\

\hline                                   
\end{tabular}
\end{table}

\begin{figure*}[htb]

\includegraphics[width=0.33\textwidth,height=0.335\textwidth]{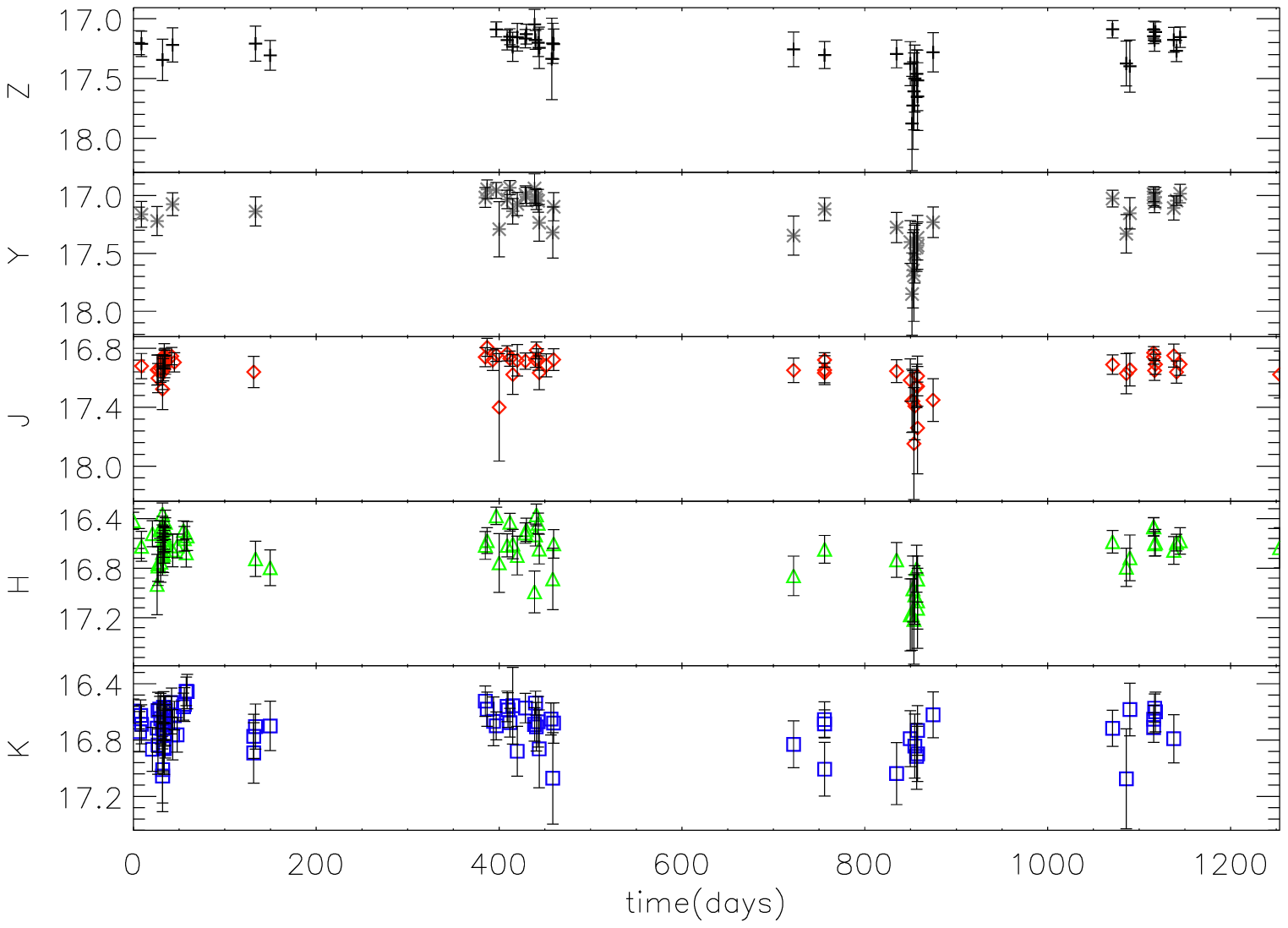}
\includegraphics[width=0.33\textwidth,height=0.335\textwidth]{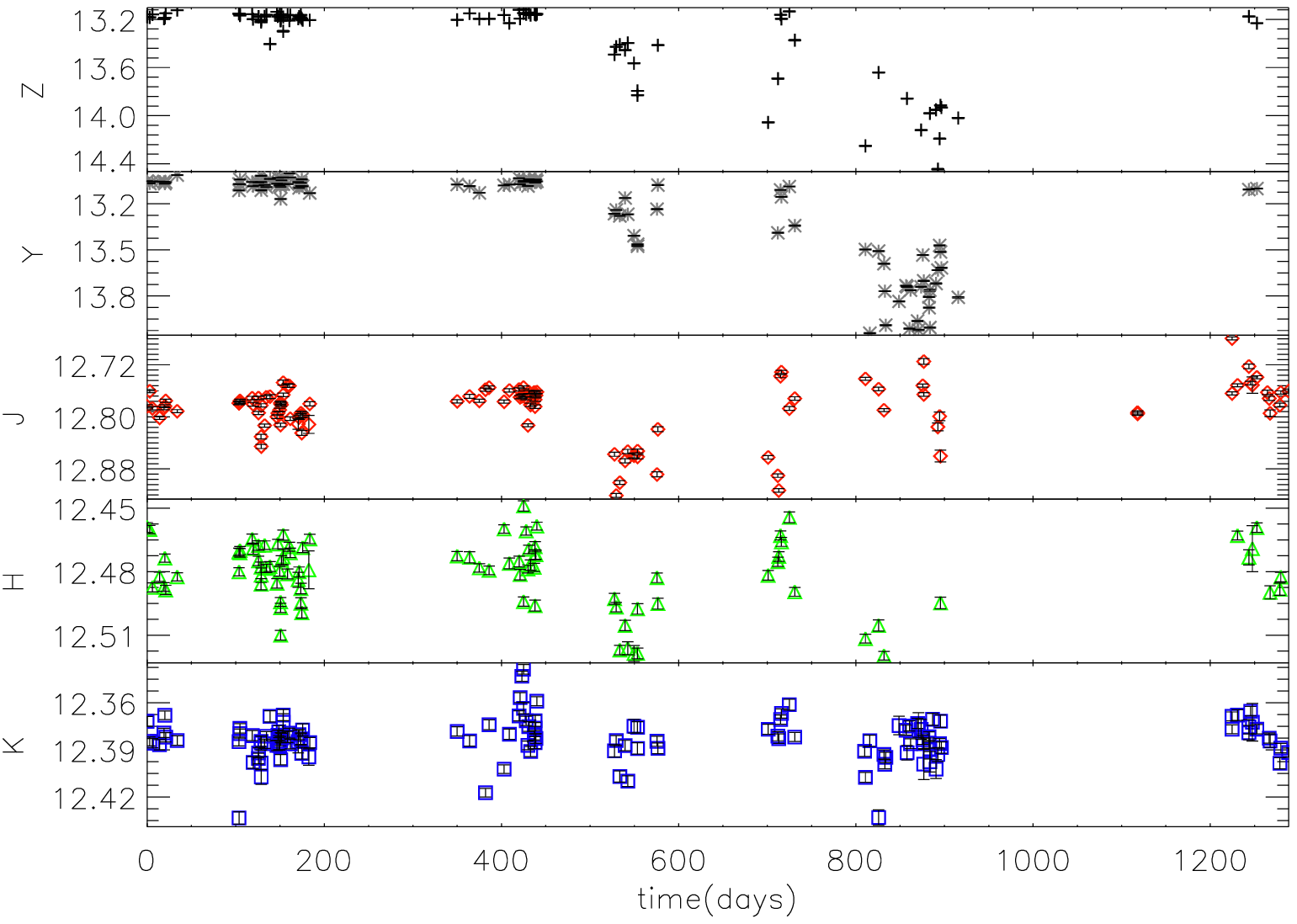}
\includegraphics[width=0.33\textwidth,height=0.335\textwidth]{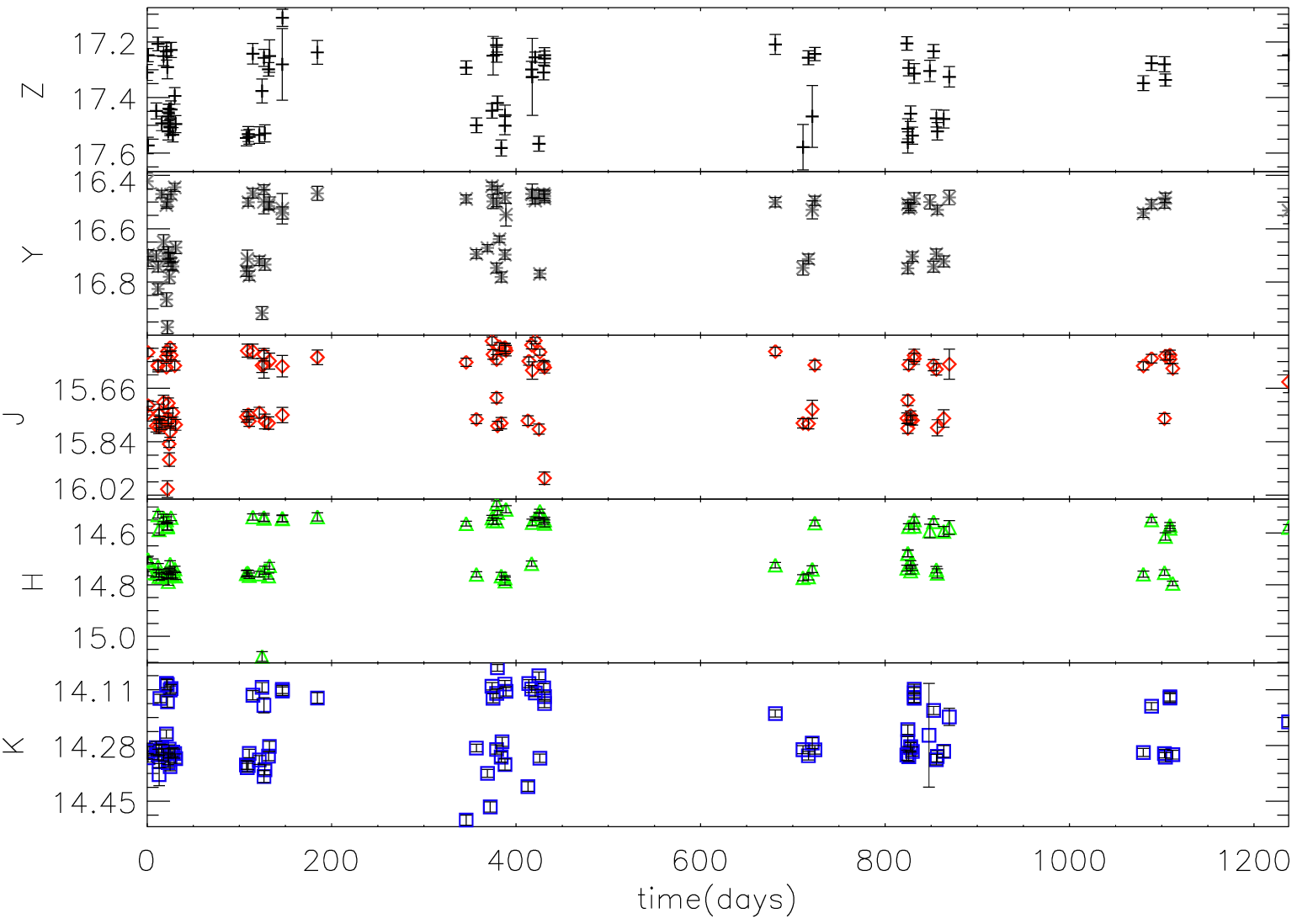}

 \caption{Examples of LCs shown intrumental bias. The identifiers are displayed
 above each panel.}
 \label{figlcvar2}
\end{figure*}

Summarizing, the correlation variability indices discriminate between uncorrelated and correlated data that is a typical feature of variable stars. We can enclose almost all WVSC1 stars in a sample with fewer than about 1500 sources. These indices still present a low efficiency for discrimination at  the faint end or bright end. The root cause in each case may well be different: excess high values for bright sources are due to temporal saturation and few epochs and increases in measurement uncertainties for the faint stars which makes the variability indices more sensitive to statistical fluctuations and systematics.

Figure~\ref{histsel} shows the histograms of sources selected for a
constant cutoff value for the higher ranked indices. These indices return
most of the WVSC1 stars with fewer non-variable stars (see Table~\ref{numsel}).
However, these indices have a clear bias on selection from the point of view of
magnitude since they are not evenly distributed along all K values. For $s = 2$
we observe that $K_{\rm fi}^{(s)}$ (black line) and
$FL^{(s)}$ (red line) display a prominent
over-selection for faint stars while $FM^{(s)}$ (blue line) is biased
for both brighter and faint stars. On the other hand, the three indices have similar bias for $s = 3$. Such a result indicates
that the efficiency of these indices may be similar for higher $s$ values since 
the difference in $E_{tot}$ between them is smaller for $s = 3$. Indeed, more
than $60\%$ of stars with $k>17.5$ have $N_{s} < 20$. This low
detection efficiency for the instrument in this region gives a maximum magnitude
limit where we can sensibly use these indices.

On the other hand, if we use the analytical expression for $f_{\rm fluc,s}$ 
(see Sect.~\ref{fap}) we obtain a higher efficiency. This function
allows us to analyse stars with lower $N_{s}$ values (of course, above the
minimum number of correlations $N_{s} > 4$) with a similar efficiency such as
that obtained for $FL^{(s)}$ and $FM^{(s)}$ considering a more strict selection, i.e. $N_{s} > 20$. Using 
$f_{fluc,s}$ we can enclose a greater number of WVSC1 stars with fewer 
contaminating sources selected (see Table~\ref{cutval}). $f_{fluc,s}$ is an 
empirical relation and it may be adapted according to its purpose.

\begin{figure*}[htb]

\includegraphics[width=0.9\textwidth,height=0.7\textwidth]{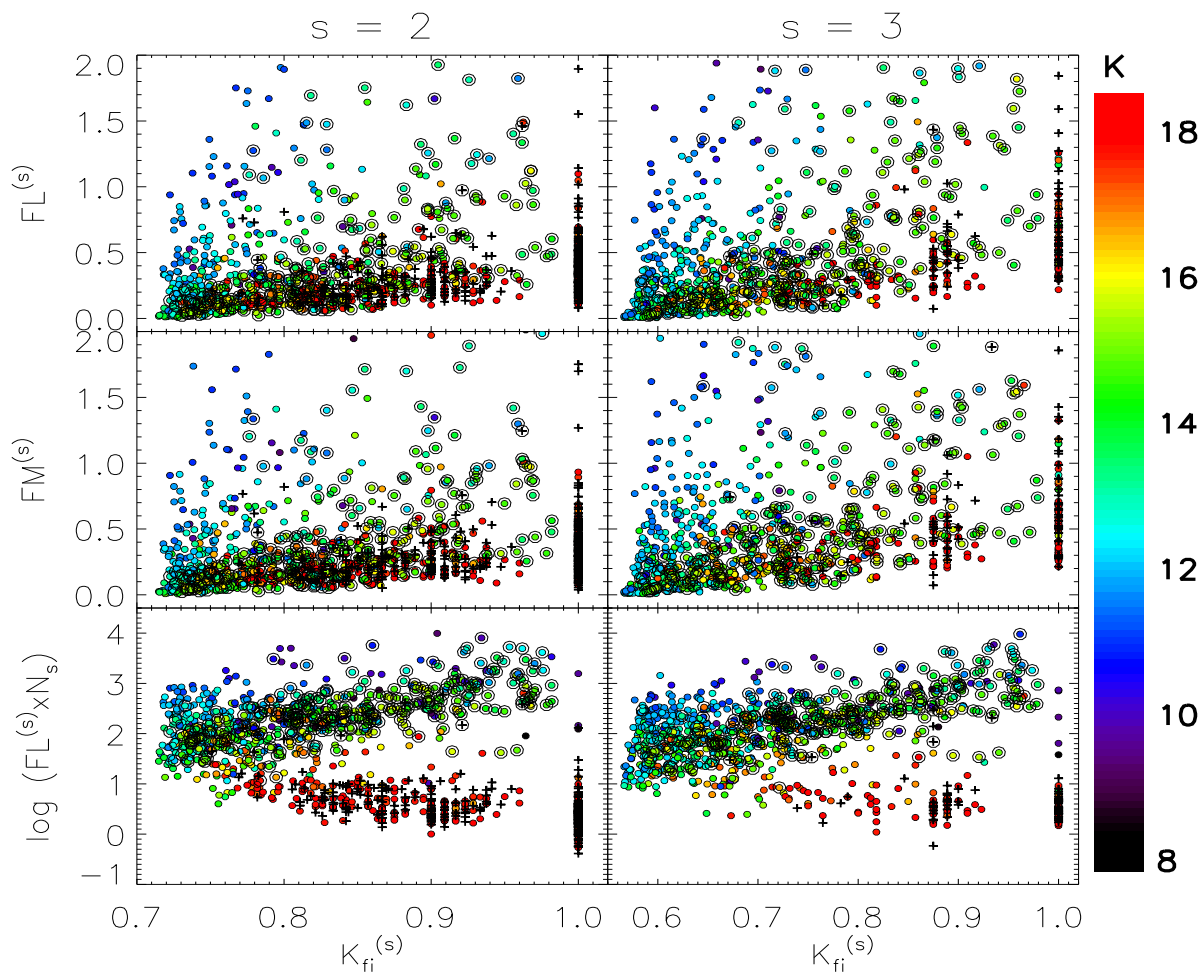}

 \caption{$FL^{(s)}$, $FM^{(s)}$, and logarithmic  $FL^{(s)} \times N_{s}$ versus the $K_{\rm fi}^{(s)}$ variability  indices for stars selected by $f_{\rm fluc}$ expression using $\alpha = 0.30$ for $s = 2$ and $\alpha = 0.46$ for $s = 3$ (see Sect.~\ref{catsel}) for $s=2$ (left panels) and $s=3$ (right  panels). The WVSC1 stars are marked by open black circles and the colours are  set by the K magnitude. Objects marked by crosses do not have a K-band  magnitude, either being too faint, or saturated. They have measurements in other bands.}
 \label{figkl}
\end{figure*}

\subsection{Searching for Variable Stars}\label{catsel}

We use the $\alpha$ values (see Sect.\ref{fap}) in order to select at least 
$90\%$ of the WVSC1 stars. Therefore, we adopt $\alpha = 0.30$ for $s = 2$ and 
$\alpha = 0.46$ for $s = 3$ which return a combined sample of $1133$ variable 
stars candidates that were not included in the WVSC1 catalogue. The periods 
were computed according to Sect.~\ref{secperiod} and we visually inspected each 
star. According to our analysis these stars can be divided in to five main 
groups: (a) variable stars measured in few epochs; (b) variable stars with low 
signal-to-noise and low confidence periods; (c) variable stars with amplitudes 
which are near to the noise level; (d) aperiodic variable stars or variables of
such long periods that these data were insufficient for deriving them; (e) false
variables due to instrumental or reduction problems.

Our procedure has resulted in a catalogue with four new sources (C1).
Fig.~\ref{figlcvar} shows the C1 stars which may be included in abc groups 
with variability indices' values near to those expected from noise. 
Fig.~\ref{figlcvar2} show some instrumental variations that can appear to give 
false positives for variable stars: false variables due to instrumental
saturation (left and middle panels) and for data reduction problems (right
panel). The separation between stars of abc and de groups is not possible using 
only variability indices. Discriminating these sources using statistical
analyses will be discussed in a forthcoming paper. Table~\ref{tab-cat01} lists 
coordinates, periods, mean magnitudes, and the number of epochs in each filter 
for this sample. WVSC-336 has a period of 192 days but its period may be higher 
since we don't observe a complete variability cycle.

\subsection{Two-dimensional View of Correlated Data}\label{bidview}

The $f_{\rm fluc}$ defined for $K_{\rm fi}^{(s)}$ variability indices, using the expression in Eq.~\ref{corrupmes}, presents the best efficiency for selecting WVSC1 stars (see Tables~\ref{cutval} and \ref{numsel}). However it returns a lot of false positives when we have few correlations. The combination 
of indices based on correlation signals ($K_{\rm fi}^{(s)}$) with those based with correlation values ($FL^{(s)}$ and $FM^{(s)}$) may provide two-dimensional view of correlated data.

Fig.~\ref{figkl} shows the $FL^{(s)}$, $FM^{(s)}$, and logarithmic of $FL^{(s)} \times N_{s}$ as a function of $K_{\rm fi}^{(s)}$ (named KFLs diagram) for $s = 2$ (left panels)  and $s = 3$
(right panels). The KFLs diagrams allow us discriminate two main groups; (G1)
faint  stars where about $90\%$ have $K_{\rm fi}^{(s)} = 1$ due to a small
number of  correlations; (G2) is composed of stars with $K \leq 16.5$ that
includes $91\%$ of WVSC1 stars. These groups display a clear separation if you 
multiply  $FL^{(s)}$ by the number of correlations ($N_{s}$) where
$G1$ has $log FL^{(s)} \times N_{(s)} < 1.5$ and G2 is delimited by
$log FL^{(s)} \times N_{s} > 1.5$. However, the last diagram is
biased for $N_{s}$ and so comparisons between different sources are
difficult to make.

Stars included in G2 with low values of $K_{\rm fi}^{(s)}$ are mainly bright, 
saturated stars showing a low level of variability and false positive variations
due to instrumental bias. The boundary between WVSC1 and other stars is not well
defined in this diagram. Nevertheless, the KFLs diagram helps us enclose about 
$90\%$ of WVSC1 stars with $E_{tot} \sim 2$ in G2.

\section{Conclusions}\label{bestsel}

\begin{table*}[htbp]
\scriptsize
 \caption[]{Periodic objects in the WFCAM Variable Star Catalog (C1).}\label{tab-cat01}
 \centering    
 \begin{tabular}{l c c c c c c c c c c c c c c c c}        
  \hline    
ID [WSA] &  ID [WVSC] & RA [deg.] & DEC [deg.] & P [d]  &  $\langle Z\rangle$ & $\langle Y\rangle$ & $\langle J\rangle$ & $\langle H\rangle$ & $\langle K\rangle$  & $N_Z$ & $N_Y$ &  $N_J$ &  $N_H$ &  $N_K$ \\
 \hline                        
858994169008  &  WVSC-335  &  +277.0394050  &  +1.7390500  &     0.40396  &  15.903  &  15.573  &  15.076  &  14.593  &  14.368  &        80  &        83  &        97  &        94  &        98 \\ 
858994205031  &  WVSC-336  &  +277.4906120  &  +1.2348900  &    192  &  17.228  &  15.128  &  12.294  &  -9.999  &  -9.999  &        68  &        11  &         8  &         0  &         0 \\ 
858994439420  &  WVSC-337  &  +104.8895590  &  -4.9367370  &    0.111042  &  12.409  &  12.250  &  12.017  &  11.822  &  11.725  &        25  &        21  &        23  &        21  &        32 \\ 
858994483642  &  WVSC-338  &  +129.0526690  &  -10.2269770  &    0.178714  &  12.531  &  12.490  &  12.275  &  12.039  &  11.984  &        11  &        10  &        11  &        11  &        11 \\ 
\hline                                   
\end{tabular}
\end{table*}

From our results we can conclude that: the analysis of databases with fewer than
4 correlated measurements is not possible using $K_{\rm  fi}^{(s)}$ and 
related indices (using factor $F^{(s)}$) when we consider, the case of one wrong
value. In these cases we may use $L_{\rm pfc}^{(s)}$ or $M_{\rm pfc}^{(s)}$ indices to
discriminate correlated and uncorrelated data. On the other hand, when we have
enough correlation measurements, the $K_{\rm  fi}^{(s)}$ variability indices 
provide unique features to do time domain analysis that allows us to define a 
general way that can be applied to any survey with correlated epochs: it
presents a low sensitive to outliers, does not undergo strong variations with
magnitude, it has a clear interpretation and a theoretical definition of a value
expected for noise, it has a well defined range of values from 0 to 1, and it
is not dependent on error bars. Therefore it may be used as a universal method 
to select correlated variations. Moreover, KFLs diagrams displays two 
unique variability features related to intensity and number of positive
correlated measurements which allow us to improve $E_{tot}$ by at least $40\%$
(see Sect.~\ref{bidview}).

The $FL^{(s)}$ and $FM^{(s)}$ values in the $KFLs$ diagrams 
(see Fig.~\ref{figkl}) may vary for different surveys. However the $K_{\rm
fi}^{(s)}$ is not strongly dependent on instrumental features and its cutoff
values  can be adopted as  universal values as can $f_{\rm fluc,s}$. Its values
are related with the discrimination of correlated and uncorrelated data and its 
response is unbiased with respect to magnitude or observed wavelength. After 
selecting the variable star candidates using $f_{\rm fluc}$ we may use the
$KFLs$ diagrams to improve the selection. Next we can remove G1 stars and use 
levels of significance of some periodicity methods to discriminate which of 
these display periodic variations.

This work is the first in a series that make a detailed analysis of all
processes of variable photometric data. In this first paper we have investigated
which indices give the most efficient selection when we have correlated 
observations. In the second paper of this series we will consider uncorrelated 
observations and determine which are the best indices for selecting variable
stars. In the coming years we will apply these methods to very large surveys of 
the Milky Way, e.g. VVV, PanSTARRS, Gaia and in the longer term to LSST to
provide fast and reliable classifications of variable stars within the Milky
Way, which will improve our understanding of the evolution of different stellar 
populations and thus the formation of structures within our Galaxy.

\section{Acknowledgements}

C. E. F. L. acknowledges a post-doctoral fellowship from the CNPq. N. J. G. C. acknowledges support from the UK Science and Technology Facilities
Council.

\bibliographystyle{aa}
\bibliography{mylib04}

\end{document}